\documentclass[twocolumn,floatfix,prb,a4]{revtex4}
\usepackage[]{amsmath}

\usepackage[psamsfonts]{amsfonts}
\usepackage[psamsfonts]{amssymb}
\usepackage{soul}
\usepackage{graphicx}
\usepackage[usenames,dvipsnames]{color}

\newcommand{\sio} {Sr$_{2}$IrO$_{4}$}
\newcommand{\lacoo} {La$_{2}$CoIrO$_{6}$}
\newcommand{\lazno} {La$_{2}$ZnIrO$_{6}$}
\newcommand{\cF} {\ensuremath{\mathcal F}}

\def\jpcm{J.\ Phys.: Condens. Matter}
\def\prl{Phys.\ Rev.\ Lett.}
\def\prb{Phys.\ Rev.\ B}

\begin{document}

\title{Study of Nontrivial Magnetism in 3$d$-5$d$ Transition Metal based Double Perovskites}
%\shorttitle{} %Insert here a short version of the title if it exceeds 70 characters

\author{Shreemoyee Ganguly}
\email[Email: ]{ganguly.shreemoyee@gmail.com}
%\shortauthor{F. Cricchio \etal}
\affiliation{Department of Basic Science and Humanities, University of Engineering and Management, University Area, Kolkata 700 160, India}

\author{Sayantika Bhowal}
\email[Email: ]{bhowals@missouri.edu}

\affiliation{School of Physical Sciences, Indian Association for the Cultivation of Science, Jadavpur, Kolkata 700 032, India}
\affiliation{Current address: Department of Physics \& Astronomy, University of Missouri, Columbia, Missouri 65211, USA}

%\pacs{71.27.+a} {Strongly correlated electron systems; heavy fermions.}

\begin{abstract} 
The coexistence of the strongly correlated 3$d$ transition metal (TM) atom and the strong spin-orbit coupling (SOC) of 5$d$ TM atom is potential to host exotic physical phenomena. In the present work, we have studied the magnetism resulting from such a coexistence in \lacoo (LCIO), a representative of bulk 3$d$-5$d$ double perovskites. In order to gain further insight into the effect of Co-$d$ states on the Ir-$d$ states, comparisons are carried out with the isostructural \lazno~, where nonmagnetic Zn atom replaces the Co atom. 
An in-depth analysis of the magnetic states in the framework of first principles calculation, using Landau theory and magnetic multipole analysis shows that the magnetism at the two constituent TM atoms in LCIO are driven by 
two different magnetic order parameters, viz., the spin moment as the primary order parameter responsible for the broken time-reversal state in Co and the higher order multipole: triakontadipole for the Ir magnetic state. A tight-binding analysis with the Ir-$t_{2g}$ orbitals,  further, indicate that the Ir-$d$ states are hardly affected by the Co-$d$ states, in agreement with the multipole analysis. The computed heirarchy of the relevant multipoles in the present work can be probed in neutron diffraction measurements, motivating further experiments in this direction.
\end{abstract}

\date{\today}


\maketitle

\section{Introduction}
5$d$ transition metal oxides (TMOs) have recieved considerable attention in recent times due to the strong spin-orbit coupling (SOC) of the heavy 5$d$ transition metal (TM) element which provides the platform to explore plethora of exotic phases ranging from  
relativistic Mott insulating state \cite{jeff} to topologically non-trivial phases \cite{Balent,IrO2,SIO1,BTIO} with possible applications in spintronics and quantum computations.
The larger extension of the 5$d$ orbitals renders a relatively weaker correlation in the 5$d$ TMOs compared to their 3$d$ counterparts. This leads to an ongoing research, attempting to combine the strong correlation of the 3$d$ TM and strong SOC of 5$d$ TM by forming 3$d$-5$d$ hetero-structures \cite{Nichols,Bhowalnpj}. Double perovskites, having the general formula B$_2$TT$^{\prime}$O$_6$ with a 3$d$ TM ion T as one of the constituent species along with the 5$d$ TM ion T$^{'}$ offer the possibility for such studies even within the bulk structure.

In the present work, we have studied magnetism in La$_2$CoIrO$_6$ (LCIO), an example of 3$d$-5$d$ double perovskite systems where both the 3$d$ (Co:$d^7$) and 5$d$ TM (Ir: $d^5$) ions are magnetically active. While the spin is a good quantum number in 3$d$ TMs, the presence of magnetic moments in $d^5$ iridates in the presence of a strong SOC i.e., without the spin as a valid quantum number, is an intriguing phenomenon. The understanding 
of this phenomenon relies on the identification of the order parameter that primarily drives the magnetism at the constituent atoms. In this context, 
the identification of the primary order parameters (POP) for the two magnetic 3$d$ and 5$d$ TM constituents can be valuable to understand the impact of the 3$d$ magnetic states on the magnetism of the 5$d$ TM atom as well.   
%This leads to the very obvious and relevant questions  are, therefore, what are the order parameters that primarily drive the magnetism in the 3$d$ and 5$d$ TM constituent ions? Are they different? If yes, then how these parameters interact with each other? 
Several experimental and theoretical studies~\cite{Kolchinskaya,Narayanan,Currie,Lee,Battle,Cao}, addressing the crystalline and the magnetic structures of LCIO, exist in the literature. A canted antiferromagnetic (AFM) structure of LCIO, where both Co and Ir sublattices are AFM, is predicted from the neutron powder diffraction (NPD) experiments~\cite{Narayanan}.
%\textcolor{red}{The structures of both the compounds La$_2$CoIrO$_6$ and La$_2$ZnIrO$_6$ are found to be monoclinic.~\cite{Battle,Lee} The Co is found to be in a high spin state [Co$^{2+}$] and Ir in low spin state [Ir$^{4+}$] in La$_2$CoIrO$_6$.~\cite{Narayanan,Lee} Neutron Powder Diffraction (NPD) based experiment shows the stable state of La$_2$CoIrO$_6$ has an antiferromagnetic magnetic structure (for the Co as well as Ir sublattice).~\cite{Narayanan} The refined moment of Co shows components of moment along x and z directions have opposite directions. However, indications are got that the y-components of Co-moments are ferromagnetcally coupled.~\cite{Narayanan} Canted antiferromagnetism is observed in La$_2$ZnIrO$_6$ through NPD experiments.~\cite{Cao}} Nevertheless, a deeper understanding of the magnetic states of La$_2$CoIrO$_6$ and La$_2$ZnIrO$_6$ and magnetic order parameters driving them is still lacking. 
However, a deeper understanding of the magnetic ground state of LCIO, in particular the role of the magnetic order parameters in driving the magnetism
in the system is still lacking.

The importance of the identification of the magnetic order parameter in the spin-orbit coupled systems like iridates lies in their unconventional magnetism, depiction of which using a model is a difficult task~\cite{SIO}. The $j_\mathrm{eff} = 1/2$ model~\cite{jeff} proposed in 2008 is the most successful model explaining the behaviour of Ir$^{4+}$. According to this model, the six-fold (including spin) degenerate $t_{2g}$ states of an Ir$^{4+}$ ion in an octahedral environment 
split into completely filled $j_\mathrm{eff} = 3/2$ quartet, and a half-filled $j_\mathrm{eff} = 1/2$ doublet, which further splits into completely filled lower and completely empty upper Hubbard band by the Coulomb interaction.  Interestingly while this model predicts a large total moment of 1 $\mu_B$ for Ir$^{4+}$ (a spin moment of 0.33 $\mu_B$ and an orbital moment of 0.67 $\mu_B$) for a broken time-reversal (TR) symmetric state, the actual value for Ir moment in \sio,  which is regarded as the archetypical ``spin-orbit Mott" insulator, is found to be much lower, both experimentally (0.208 $\mu_B$/Ir)~\cite{Ye} as well as 
from the first principle calculations (a spin moment of 0.08 $\mu_B$/Ir and an orbital moment of 0.24 $\mu_B$/Ir)~\cite{Hongbin,SIO}. This discrepancy, as explained in an earlier work ~\cite{SIO}, by one of us, follows from the fact that the source of the broken TR symmetry in \sio, is a higher order magnetic multipole moment originating from the entanglement of spin and orbital 
magnetic orders~\cite{SIO}. The ordered spin magnetic moments are found to be merely secondary order parameter (SOP) induced by the higher order multipole which happens to be the POP. 

In view of this, it is important to identify the POP for the individual 3$d$ and 5$d$ species Co and Ir in LCIO which, in turn, can elucidate the influence of the these two magnetic species on each other. In the present work, fixed spin moment calculation in the light of Landau theory and explicit multipole analysis of LCIO are carried out to understand the relative hierarchy 
and the roles of the different magnetic order parameters in driving the magnetic state in the system. Our calculation shows that LCIO is a striking example, where the magnetism in the two different species are driven by the two different order parameters, viz., spin moment at the Co site and the  higher-order multipole of rank five (triakontadipole) at the Ir site.  
In order to gain further insight into the role of the Co ion on the magnetism of the Ir ion, we have also studied and compared the magnetism with the isostructural double perovskite iridate La$_2$ZnIrO$_6$ (LZIO), where Zn$^{2+}$ is magnetically inactive in contrast to Co in LCIO. The results of the multipole analysis are  corroborated by our tight-binding model analysis, addressing the impact of the Co-$d$ states on the Ir-$d$ states.
%relevance of the $j_{\rm eff}$ = 1/2 state in LCIO.

The remainder of the paper are organized as follows. A detailed description of the computational techniques and multipole analysis are given in section II, followed by the structural details of the double perovskite iridates LCIO and LZIO in section III. The results of our calculations are discussed in section IV. In this section, we have first analyzed and compared the basic electronic structure, viz., crystal field splittings, TM-O covalencies, and the magnetic ground states of LCIO and LZIO. This is followed by the fixed spin moment calculations, the results of which are illustrated using extended Landau free energy analysis, and the magnetic multipole analysis to determine the POP for Co and Ir magnetic states. The influence of the Co-$d$ states on the Ir-$d$ states is further studied within the tight-binding analysis. Finally  a summary of our work is presented in Sec. V.

\section{Computational Details}

In order to investigate the hierarchy of the different magnetic multipoles of LCIO, first the electronic structure is obtained through first principles calculation based on density functional theory (DFT) with the local density approximation (LDA). In this study, we have used the augmented plane wave plus local orbitals (APW+{\em lo}) method as implemented in the {\sc ELK} code~\cite{elk}. Self consistency is achieved within DFT+SOC+U approach with a k-mesh 6$\times$6$\times$4 , where the SOC is treated second variationally.
The Coulomb correlation parameter U is chosen to be 5 eV at the Co site in LCIO and  3 eV at the Ir site in both LCIO and LZIO in accordance with trends 
followed in literature \cite{Narayanan, SIO}. With these choices of the U parameters, a reasonable description of the magnetic properties of LCIO and LZIO can be obtained. In this approach for a given value of U, the intra-atomic exchange coupling energy (J$_{\rm H}$) is calculated using a screened Coulomb potential (Yukawa potential), as implemented in the code 
{\sc Elk}~\cite{LAPW,Bultmark-Mult}. This approach calculates the ratio of the Slater integrals in a much more accurate way than other 
implementaions of DFT+U approach where both U and J$_{\rm H}$ need to be specified. The computed values of J$_{\rm H}$ are 1.03 eV for Co and 0.94 eV for Ir. The localized limit was adopted for the double counting correction. 

\begin{figure}[t]
\centering
\includegraphics[width=\columnwidth]{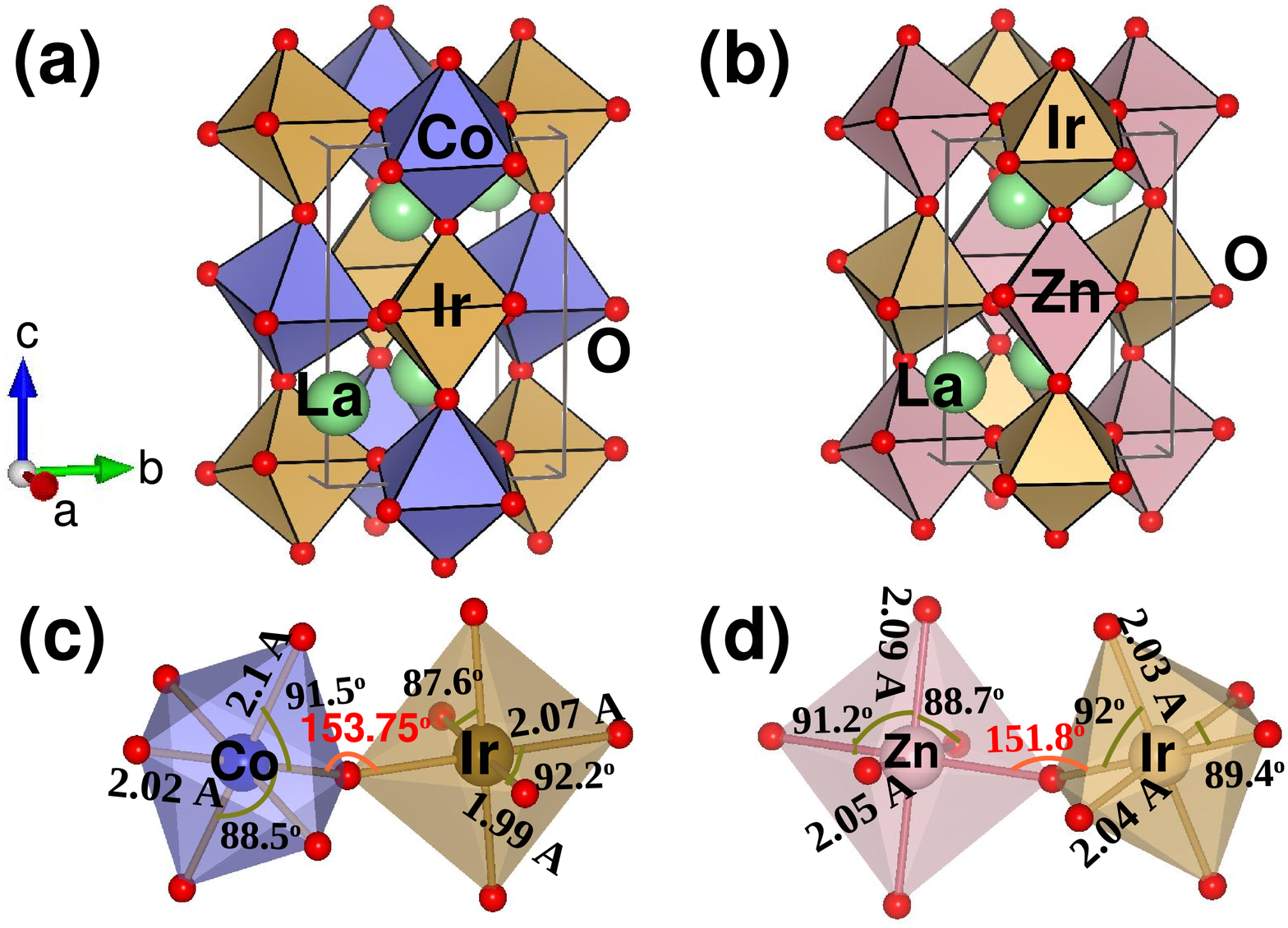}
\includegraphics[width=\columnwidth]{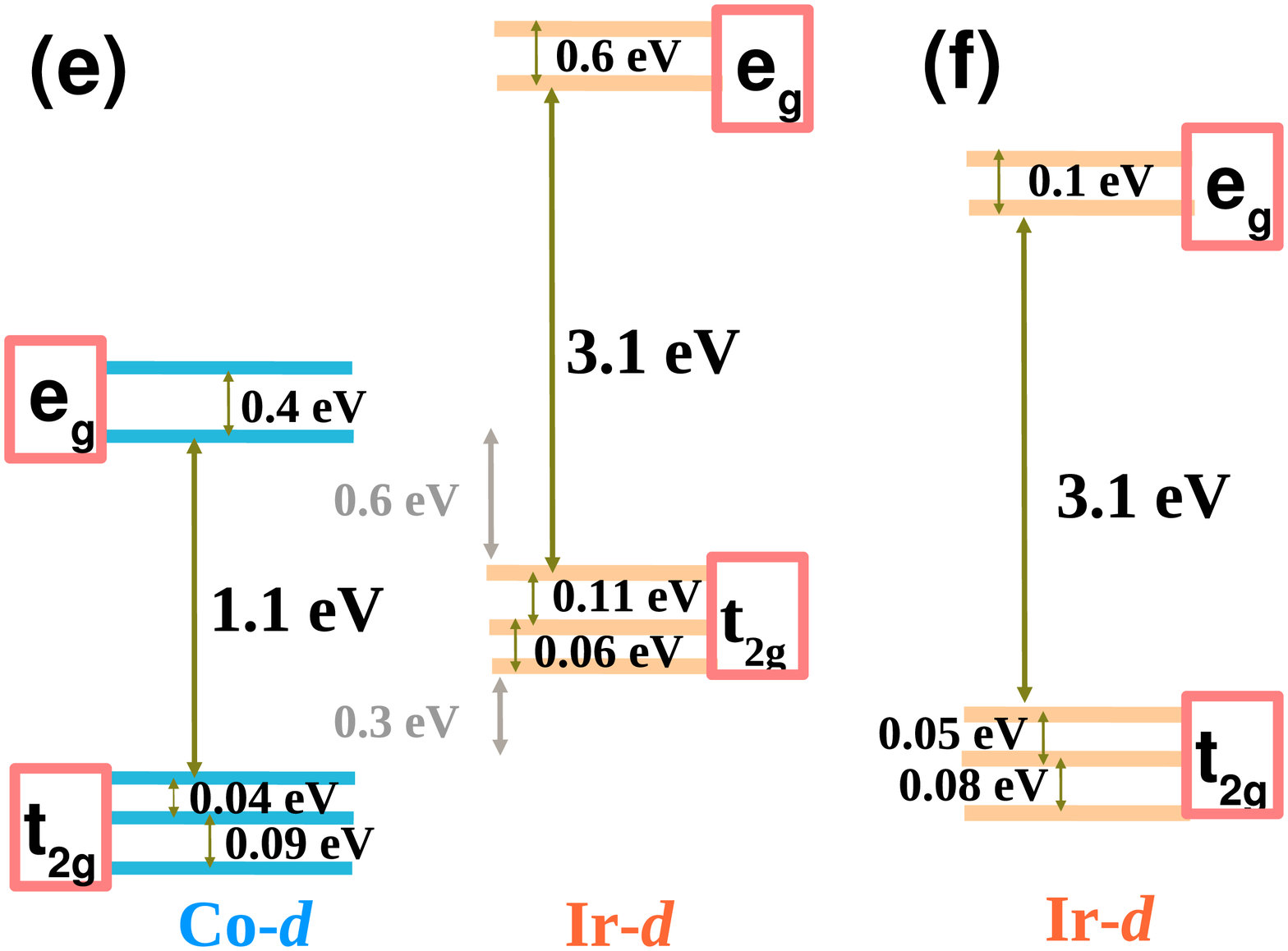}
\caption{Structural details and crystal field splitting. The monoclinic crystal structures of (a) LCIO and (b) LZIO. The distorted IrO$_6$ and MO$_6$ (M = Co, Zn) octahedra in the respective structures are shown in (c) and (d). The 
various bond angles and bond lengths of the octahedra are also indicated. The resulting crystal field splittings are illustrated for the (e) Co-$d$ and Ir-$d$ states in LCIO and (f) Ir-$d$ states in LZIO.} 
\label{fig1}
\end{figure}
{\it Multipole analysis:} In order to carry out the multipole analysis of the obtained stable solutions, we expand 
the rotationally invariant local Coulomb interaction in terms of the multipole tensors 
in the mean field limit \cite{Bultmark-Mult}, a brief description of which is given in the Appendix. In the present study, with Co-$d$ and Ir-$d$ orbitals, the density matrix $D$ ($D$ = $<d_nd_n^{\dagger}>$) has 100 independent 
elements. Here $d_n^{\dagger}$ and $d_n$ are the creation and the annhilation  operators 
respectively for the Ir-$d$ and Co-$d$ states. This information, carried 
by the density matrix, can be transformed to the expectation values of the multipole 
tensor moments $w^{kpr}$ = $Tr\Gamma^{kpr}D$, where the multipole   
tensor operator $\Gamma^{kpr}$ is an Hermitian matrix operator~\cite{Cricchio}. The indices are determined by coupling of the spin and the  orbital angular momenta, eg., : 0~$\leq~k~\leq~2l$ = 4, 
0~$\leq~p~\leq~2s$ = 1, and $|k-p|$~$\leq~r~\leq$~$|k+p|$. There are 18 such multipole tensors $w^{kpr}$ 
which together have 100 tensor components, corresponding to $l = 2$ ($d$-states). Out of these 18 multipole tensors, only 9 of them break the TR symmetry. These tensor moments have very nice 
correspondance to the physical entities charge, magnetization and spin current. The moments $w^{kpr}$ that correspond to {$k$ even and $p = 0$}, {$k$ even and $p = 1$}, {$k$ odd and $p = 0$}, and {$k$ odd and $p = 1$} are 
proportional to moment expansions of the charge, magnetization, current and spin-current respectively. Amongst these only the 
moments associated with magnetization and current break the TR symmetry and are thus considered in our present work. 

Further, to have a quantitative estimate of the crystal field splitting and hopping parameters, we have employed the muffin-tin orbital (MTO) based N$^{th}$ order MTO (NMTO)~\cite{NMTO1, NMTO2, NMTO3} downfolding method as implemented in the Stuttgart code keeping either both Ir-$d$ and Co-$d$ states or only Ir-$d$ states in the basis, integrating out the high energy degrees of freedom.
\section{structural details}
Both LCIO and LZIO crystallize in the monoclinic structure with the space group $P2_1/n$~\cite{Battle,Lee}. The crystal structures consist
of corner sharing IrO$_6$ and MO$_6$ (M = Co, Zn) octahedra, alternating along each of the crystallographic axes as shown in Fig~\ref{fig1}(a) 
and (b) respectively. The unit cell contains two formula units, which means two Ir atoms and two M (= Co, Zn) atoms are present in the unit cell of LCIO and LZIO.
The monoclinic symmetry of the structures allows the rotation of the IrO$_6$ octahedra with $\angle$Ir-O-M $\sim$ 154$^\circ$ and 
$\sim$ 152$^\circ$ for LCIO~and LZIO ~[See Fig~\ref{fig1}(c) and (d)] respectively. The Ir-O bond lengths are also not equal for both the iridates, and the variation is larger in LCIO~($\sim$1.99-2.07 \AA) compared to LZIO~($\sim$2.03-2.04 \AA). The distortion of the IrO$_6$ 
octahedra splits the Ir-$t_{2g}$ states completely into three non-degenerate states as will be discussed in the next section.

\section{Results and Discussions}

\subsection{Crystal field splitting and covalency}
\begin{figure}[t]
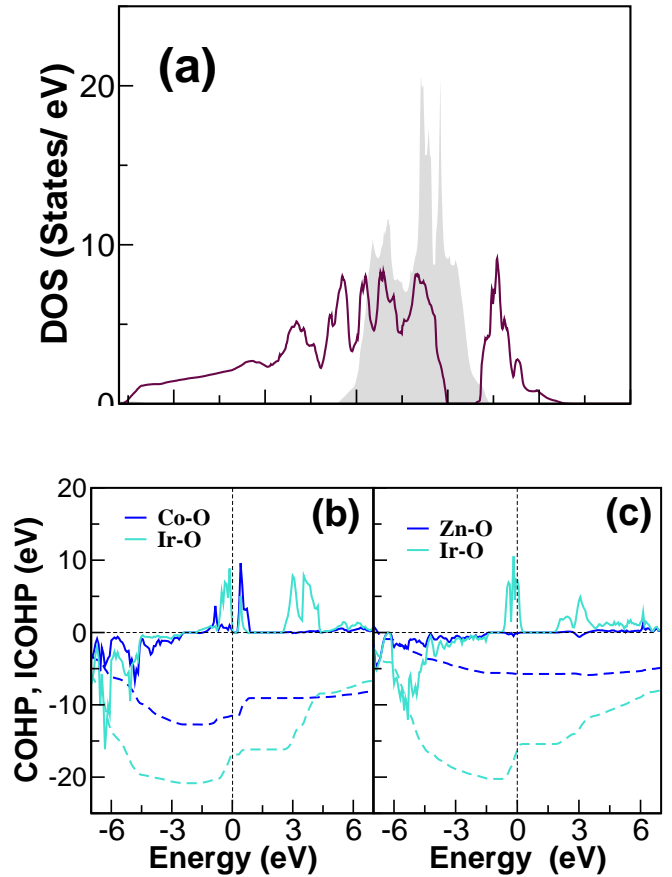

\centering
%\vspace{2 cm}
\includegraphics[scale=0.35]{fig2a.eps}
\includegraphics[width=\columnwidth]{fig2b.eps}
\caption{ Band-width and covalency. (a) Comparison of  Ir-$d$ projected densities of states (DOS) in LCIO (solid line) and  LZIO (shaded region) in absence of magnetism and SOC, indicating a smaller band-width in LZIO compared to LCIO. COHPs (solid line) and integrated COHPs (ICOHP) (dotted line) for the M-O (M = Co, Zn) and the  Ir-O bonds in (b) LCIO and (c) LZIO. A stronger Ir-O covalency compared to Co (Zn)-O is evident from these plots, as expected.}
\label{fig2}
\end{figure}
{\it Crystal field splitting:} As we have discussed earlier, both Co and Ir atoms are in the octahedral network. In absence of magnetism and SOC, an estimation of the crystal field splittings at the Co-$d$ and Ir-$d$ states in LCIO [see Fig~\ref{fig1} (e)], shows that the $t_{2g}$-$e_{g}$ splitting is stronger at the Ir site ($\sim$ 3.1 eV) than the Co site ($\sim$ 1.1 eV), as expected due to the extended 5$d$ orbitals of Ir. From the energetics given in Fig~\ref{fig1} (e), it is clear that the Ir-$t_{2g}$ states are lying within the energy gap of Co-$t_{2g}$ and Co-$e_g$ states, viz., at an energy $\sim$ 0.3 eV higher than the Co-$t_{2g}$ states and $\sim$ 0.6 eV below the Co-$e_g$ states. 
For comparison the crystal field splitting at the Ir-$d$ states in LZIO is depicted in Fig~\ref{fig1} (f). Although the $t_{2g}$-$e_g$ splitting is found to be similar to LCIO, the smaller distortion of the IrO$_6$ octahedra in LZIO leads to a weaker non-cubic crystal field compared to LCIO. 

Further, the plot of the densities of states (DOS), depicted in Fig~\ref{fig2} (a), clearly shows that the band-width is narrower in LZIO compared to LCIO. This may be attributed to the presence of the Co-$d$ states in LCIO that hybrize with the Ir-$d$ states via the oxygen atoms as discussed later. The smaller band-width and weaker noncubic crystal field 
in LZIO compared to LCIO indicates a stronger effective SOC in LZIO than LCIO. 
%This is also reflected in the multipole analysis described in section \ref{multipole}.

%Now, we introduce magnetism in the system and study its band gap. The band gap of \lacoo~as a function of the enhanced correlation $U$ at Co site is listed in Table \ref{bndgap}.  

%\begin{table}[h]
%\caption{Variation of band gap of \lacoo~with the correlation $U$ at Co site.}
%\centering 
%\begin{tabular} {|l|l|l|}
%\hline 
%U$_{Co}$(in eV) & U$_{Ir}$(in eV) & Band Gap (in meV)\\
%\hline
%4        & 3 & 570.31 \\
%5        & 3 & 590.23 \\
%6        & 3 & 600.47\\
%\hline
%\end{tabular}
%\label{bndgap}
%\end{table}
%Keeping $U$ at Ir site fixed (U$_{Ir}$=3 eV), as we increase the correlation at Co site U$_{Co}$ from 4 eV to 6 eV (as seen from Table \ref{bndgap}) the band gap of \lacoo~increases from 570.31 meV to 600.47 meV. The band gap of \lazno~with enhanced correlation at Ir site U$_{Ir}$ set to 3 eV is 544.23 meV.

{\it Covalency:} In order to get a quantitative understanding of the strength of covalency of the Ir-O and M-O (M = Co and Zn for LCIO and LZIO respectively) bonds, we  have computed the crystal orbital Hamiltonian population (COHP) as implemented in the Stuttgart
tight-binding linear muffin-tin orbital (TB-LMTO) code.~\cite{AndersenJepsen} The COHP and the integrated COHP (ICOHP) give the information of the nature of a specific bond between a pair of atoms and the integrated value of the strength of such interactions respectively. 
The results of our calculation which are presented in Fig.~\ref{fig2} (b) and (c), show the off-site COHP and the energy integrated COHP (ICOHP) per bond for the nearest neighbor Ir-O and M(= Co, Zn)-O in LCIO and LZIO respectively.
These COHP plots represent the energy resolved visualization of the chemical bonding between Ir-O atoms, and M (M = Co, Zn) -O atoms. In COHP, the DOS is
weighted by the Hamiltonian matrix elements. The off-site COHP represents the covalent contribution to the bonds.~\cite{cohp} In Fig~\ref{fig2} (c) and (d), the bonding contribution for which the system undergoes a lowering in energy is represented by a negative value of COHP, while the antibonding contribution, which raises 
the energy, is indicated by the positive value. This gives a quantitative measure of bonding. 
It is clear from Fig.~\ref{fig2} (b) and (c) that the Ir-O covalency is substantially stronger compared to Co/Zn-O covalency. This can be understood from the extended nature of the 5$d$ orbitals compared to the 3$d$ orbitals of Co or Zn. Also, the Co-O hybridization is much stronger than the Zn-O covalency. 
%This results in slightly stronger Ir-O covalency in LCIO compared to LZIO, consistent with the larger band-width in LCIO. 
We note that, due to large Ir-Co distance the direct hybridization between Ir-Co/Zn is weak resulting in a negligibly small Ir-Co covalency. However, the significant ICOHP corresponding to the Ir-O and Co-O bonds indicate the existence of super-exchange interactions between Ir and Co ions, mediated by the oxygen atoms. 

\subsection{Ground state Magnetic structure}

In order to figure out the magnetic order parameters, we have first analyzed the ground state magnetic structures of both LCIO~\cite{Narayanan} and LZIO\cite{Cao}. Both these isostructural double perovskite iridates with $P2_1/n$ crystallographic symmetry are predicted to be 
compatible with the propagation vector {\bf k} = (0, 0, 0) \cite{Cao}. According to the symmetry analysis, the given propagation vector allows four possible magnetic
structures \cite{Bilbao}, viz., P2$_1^{'}$/C$^{'}$, P2$_1$/C, P2$_1$/C$^{'}$, and P2$_1^{'}$/C, out of which P2$_1$/C$^{'}$, and P2$_1^{'}$/C correspond to the nonmagnetic structure. 

\begin{figure}[t]
\centering
%\vspace{2 cm}
%\resizebox{12cm}{!}
{\includegraphics[width=\columnwidth]{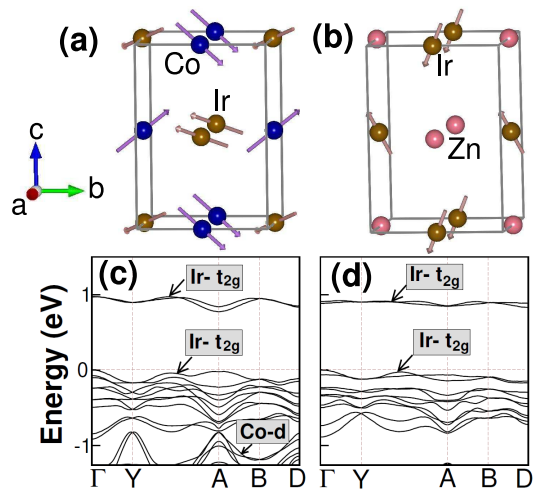}}
%
%{\includegraphics[width=\columnwidth]{fig3b}}
%\includegraphics[scale=.5]{fig6.eps}
\caption{The ground state magnetic structures, computed from DFT calculation, for (a) LCIO and (b) LZIO. The arrows denote the directions of the spins at the magnetic sites in both cases. The corresponding band structures for (c) LCIO and (d) LZIO. The Fermi energy is set at zero. The predominant orbital characters of the bands (computed from the explicit projection of band characters corresponding to various atomic species) are indicated in the figure.}
\label{LCIO-SOC}
\end{figure} 
\begin{table}[t]
\caption{ Comparison of the total energies computed within DFT+SOC+U for various symmetry allowed magnetic configurations (see text for details) of LCIO and LZIO. The total energy of  $P2_1/C$ is set at zero for both the iridates. The listed energy differences are in units of meV/atom for LCIO, while for LZIO the values are in meV/f.u. }
\centering
\begin{tabular}{c c c }
\hline
 & \multicolumn{2}{c}{$\Delta$E for}\\[0.3 ex]
\hline
Configuration\ &  LCIO \ &  LZIO \\[0.3 ex]
              \ &    (meV/atom) \ & (meV/fu) \\
\hline
 NM      &     140       &  198\\
 ($P2_1/C^\prime$,$P2_1^\prime/C$) &   &    \\
 $P2_1^\prime/C^\prime$     & 10    &  2 \\
 $P2_1/C$                   &  0    & 0   \\
\hline
 \end{tabular}
\label{tab}
\end{table}
Calculations in presence of SOC and Coulomb correlation U for LCIO 
and LZIO show that the magnetic ground states, as depicted in Fig. \ref{LCIO-SOC} (a) and (b), correspond to the 
P2$_1$/C magnetic space group. The energy differences among the different magnetic structures are listed in Table \ref{tab}. In the computed non-collinear magnetic ground state, the $x$ and $z$ components of the Ir and Co moments are antiferromagnetically aligned with each other while their $y$ components are parallel. The calculated values of the moments at the two Co ions 
in LCIO are: $\mu_x$~=~1.12$\mu_B$, $\mu_y$~=~1.84$\mu_B$, and $\mu_z$~=~-1.02$\mu_B$; and 
$\mu_x$~=~-1.12$\mu_B$, $\mu_y$~=~1.84$\mu_B$, and $\mu_z$~=~1.02$\mu_B$. Here, $\mu_x$, $\mu_y$, and $\mu_z$ 
denote the spin moments along the $x$, $y$, and $z$-directions respectively. The spin moments at the two Ir sites are: $\mu_x$~=~-0.04 $\mu_B$, $\mu_y$~=~-0.04 $\mu_B$, and $\mu_z$~=~-0.01 $\mu_B$; and $\mu_x$~=~0.04$\mu_B$, 
$\mu_y$~=~-0.04 $\mu_B$, and $\mu_z$~=~0.01 $\mu_B$. 
 This is in accordance with the magnetic structure found in neutron powder diffraction (NPD) measurements \cite{Narayanan} for LCIO.
 The corresponding moments at the Ir sites in LZIO are found to be: $\mu_x$~=~-0.03$\mu_B$, 
$\mu_y$~=~-0.05$\mu_B$, and $\mu_z$~=~-0.07$\mu_B$; and $\mu_x$~=~0.03$\mu_B$, $\mu_y$~=~-0.05$\mu_B$, and 
$\mu_z$~=~0.07$\mu_B$. The obtained canted antiferromagnetic structure for LZIO is also in agreement with the NPD experiments~\cite{Cao}.
 The band structures corresponding to the computed magnetic ground states of LCIO and LZIO are shown in Fig. \ref{LCIO-SOC} (c) and (d) respectively. As seen from these figures, the valence bands and the conduction bands near the Fermi energy are predominantly formed by the occupied and the unoccupied Ir-$t_{2g}$ states ($t_{2g}^5$) in both the iridates.
\subsection{Fixed Spin Moment Analysis}

In order to determine whether the spin magnetic moment of the Co atom is the dominant POP for the magnetism at the Co site in LCIO, we have  
carried out fixed spin moment analysis for the said compound. 

\begin{figure}[ht]
\centering
\includegraphics[scale=0.35]{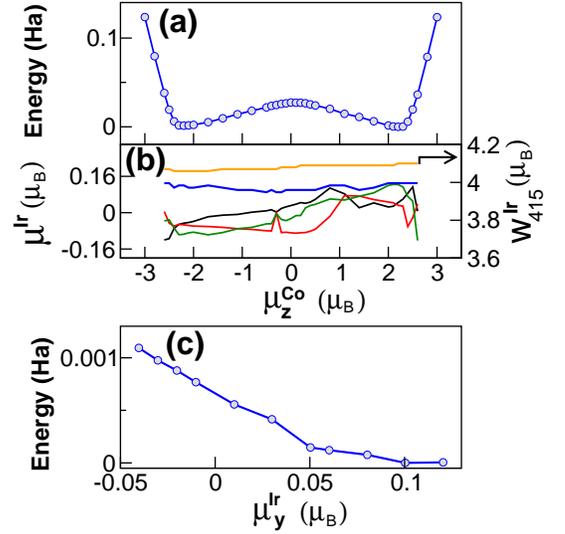}
\caption{Results of the fixed spin moment calculations. (a) The total energy as a function of the constrained spin magnetic moment component $\mu_{z}^{Co}$ (setting $\mu_{y}^{Co}$ to zero) at the Co-site of LCIO shows a quadratic behaviour about the equilibrium value  
%{\color{red}At this value of $\mu_{z}^{Co}$ the value of the total energy of the system is lowest. So, this is the value of spin moment that the system would take spontaneously. But in this case this lowest energy moment $\mu_{z}^{Co}$ is got under the constrain $\mu_{y}^{Co}$ =0. So, although it is the lowest energy point in this analysis it is not the spontaneous value of z-component of moment.}{\color{blue} [Shreemoyee di: Why is it called equilibrium value?]} 
of the moment. The point $\mu_{z}^{Co}$ = 0 is a special point as the energy exhibits mirror symmetry about it, i.e., the energy remains the same as $\mu_{z}^{Co}$~$\rightarrow$~-$\mu_{z}^{Co}$ (see the extended Landau free energy analysis in the text for details). (b) The variation of the different components of the Ir spin moments [$\mu_x^{Ir}$ (black), $\mu_y^{Ir}$ (red), and $\mu_z^{Ir}$ (green)], the total spin moment [$\mu_{\rm tot}^{Ir}$ (blue)], and the Ir triakontadipole moment [$w^{\rm Ir}_{415}$ (orange)] as a function of the $z$-component of the Co moment ($\mu_{z}^{Co}$). Although the individual components of the spin moment vary, 
%$\mu_x$, $\mu_y$ and $\mu_z$ change with the variation of $m_{z}^{Co}$, 
the total spin moment $\mu_{\rm tot}^{\rm Ir}$ and the triakontadipole moment $w^{\rm Ir}_{415}$ 
remains more or less constant.  (c) The energy variation in LCIO as a function of the variation of $\mu_y^{Ir}$ (suppressing the dominant Co moment), which resembles one of the parabolas described in Eq. \ref{fixspin2}. The energy variation is  significantly different from that in (a), indicating that the spin moment is not the POP for the magnetism at the Ir site.} 
\label{fig3}
\end{figure}

 In this calculation, we have fixed $z$-component of the Co spin moment, without imposing any constrain on the $x$-component of the moment, which is self-consistently converged to an optimal value. Note that the $z$-components of the spin moments at the two Co-sublattices are kept anti-parallel to each other. 
%one of the spin components of the Co moment, keeping the non-collinear magnetic structure of LCIO. Since it is good enough to vary any component of the moment, in the present analysis we choose to vary the exprimentally accessible $z$-component. 
Further, for the ease of the computation, we have suppressed the $y$-component of the Co-moment, and studied the nature of the corresponding variation in the total energy of the system. The $z$-component of the local Co spin moments $\mu_{z}^{Co}(\vec{R}_n)$ are 
constrained \cite{Dederichs} by auxiliary constraining fields $h_{s,z}(\vec{R}_n)$, where $n$ runs over the Co atoms (with volume $S_n$),
\begin{widetext}
%\begin{equation}
\begin{align} 
E\left(\mu_{z}^{Co}\right)=\min \left\{ E_{DFT+U} + \sum_n h_{s,z}(\vec{R}_n) \times \left(\int_{S_n} \hat{z}\cdot \vec{\mu}^{Co}(\vec{r})\mathrm{d}V - \mu_{z}^{Co}(\vec{R}_n)\right)\right\},\,\label{fixspin}
\end{align}
%\end{equation}
\end{widetext}

%\begin{widetext}
%\begin{equation*}
 %E (\mu_{z}^{Co})= \min \{ E_{DFT+U} + \sum_n h_{s,z}(\vec{R}_n) \times (\int_{S_n} \hat{z}\cdot \vec{\mu}^{Co}(\vec{r})\mathrm{d}V - \mu_{z}^{Co}(\vec{R}_n))\},\,\label{fixspin}
%\end{equation*}
%\end{widetext}

under the condition that $\mu_{y}^{Co}(\vec{R}_n)$ = 0.
%{\color{red}Actual Reason why y component of the Co moment was not constrained: It was not possible to have a controlled variation of the y-component. Reason we might put forward to the referee: it is good enough to vary any component of moment. We choose to vary the exprimentally accessible z-component instead of the y-component which is not experimentally detectable even by NPD. We may also put forward the actual reason as well, if questioned. 
The terms within the parenthesis in Eq. \ref{fixspin} gives the difference between the $z$-component of the magnetic moment that the Co atom has at a certain step of optimization and the $z$-component of the spin 
magnetic moment that we want the Co atom to have. 
A constraining field proportional to this 
difference is applied to force the $z$-component of the Co moment to the desired value. The variation of the corresponding energy is shown in Fig. \ref{fig3}.

We find from Fig. \ref{fig3} (a), that the total energy of the system possesses mirror symmetry about the point $\mu_{z}^{Co}$ = 0, i.e., if $\mu_{z}^{Co}$~$\rightarrow$~-$\mu_{z}^{Co}$, the total energy remains the same. This shows that the Co spin moment is a dominant magnetic order parameter, viz., the POP in the system. If it had been merely a SOP, the total energy of the system would not be mirror symmetric as a function 
of its variation, as can be seen from the extended Landau free energy theory, explained in details in the latter part of this section.
%.}{\color{blue} [Shreemoyee di: Why the mirror symmetry in energy indicates that the Co spin moment is the POP? What is the argument from the extended Landau free energy theory? May be we can discuss this point in more detail to enhance the readability.]} 

In order to see the effect of the Co spin moments on the spin moments of the Ir, we have further studied the variation of the components of the Ir spin moment as a function of the variation of the $z$-component of the Co moment, as discussed above. 
The resulting variation in Fig. \ref{fig3} (b)
shows that although $x$, $y$, and $z$-components of the Ir moments change as we vary the $z$-component of the Co moment ($\mu_{z}^{Co}$), the total Ir moment remains almost constant. The rotation of the Ir moment as we change $\mu_{z}^{Co}$, keeping the magnitude of the Ir moment unchanged is  
indicative of the fact that the Ir spin moment is not completely driven by the Co spin moment although the latter is a dominant POP 
for the Co magnetic state. Further, to see the effect of the POP of Co on the POP of Ir (which is the triakontadipole moment $w^{\rm Ir}_{415}$, as discussed latter in Table \ref{tab1}), we also studied the corresponding variation which is depicted in Fig. \ref{fig3} (b). As we can see from Fig. \ref{fig3} (b), the magnitude of the triakontadipole moment $w^{\rm Ir}_{415}$ more or less remains constant with the variation of $\mu_{z}^{Co}$, suggesting that the POPs of Co and Ir do not affect each other significantly.

%{\color{red}It is not possible to do constrained variation of $w^{415}$ using ELK as yet. Yes, if we could we should have got the mexican hat variation. Yes, I agree it is the POP for Co magnetic state. But it should also be a dominant OP for the system considering the high exchange energy associated with it. The $w^{415}_{Co}$ also has a high Exchange energy associated with it. But we must remember $w^{415}_{Co}$ is driven by $w^{011}_{Co}$ and does not have independent existence.}{\color{blue} [Shreemoyee di: Is it really the POP for the system or the POP, responsible for the magnetism in Co? What happens if we vary $w^{415}$ of Ir? Are not we expect the similar mexican hat in energy? Am I missing something?]}, and the POP for the Co magnetic state. 

Similar to the case of Co, we also did the fixed spin moment analysis for the Ir spin moment. For this study we suppress the dominant Co moment to be zero.  In this case, we have fixed the $y$-component of the Ir spin moment, while the $x$ and the $z$-components of the moment are optimized in a self-consistent calculation.  
%{\color{red}It is possible to do constrained variation of one component of spin moment at a time. The antiferromagnetic z-component was varied, the large y-component was interfering and not letting the z-component to be varied smoothly, so it was set to zero. Doing this exercise we do get the correct behaviour of energetics with variation of one of the components of the POP for Co magnetic state. So the exercise should not be unphysical-right?}{\color{blue} [Shreemoyee di: Why do we need to suppress the Co moment? Is it physical?]}. The corresponding evolution of the energetics of LCIO is depicted in Fig. \ref{fig3} (c).  
It is clear from Fig. \ref{fig3} (c) that the energy evolution with the variation of the $y$-component of the Ir moment ($\mu_{y}^{Ir}$) shows a very different behaviour from its evolution with Co moment depicted in Fig.~\ref{fig3} (a). The point $\mu_{y}^{Ir}$ = 0 is no longer a special point as the energy is not symmetric around that point. This behaviour is quite similar to the behaviour seen in \sio~\cite{SIO}. Based on the analysis of the extended Landau free energy theory, which we proceed to discuss next, it is possible to show  that the asymmetry in the energy with the variation of the Ir spin moment originates from the fact that 
the said spin moment is just a SOP of the Ir magnetic state. 

%in particular and of the system under study in general.

{\it Extended Landau free energy analysis:} 
The results of the fixed spin moment calculations, as discussed above, can be understood in the light of the extended Landau free energy analysis.
%As the OPs of one magnetic sublattice do not seem to affect the other sublattice significantly [see Fig.~\ref{fig3} (b)], 
 In the present analysis, for simplicity we consider the 
magnetic OPs of Co and Ir sublattices in LCIO to be decoupled. Such assumption may be justified as the POPs of the two sublattices do not seem to affect each other significantly [see Fig.~\ref{fig3} (b)]. For the Co sublattice the spin moment $\mu_{Co}$ is the 
POP. Thus the Landau free energy ($\cF_{Co}[\mu,w]$ ) of this sublattice in terms of the two independent weakly interacting TR odd OP spin moment 
$\mu_{Co}$ and the triakontadipole moment $w_{Co}$ (which is the second most significant TR odd OP as can be seen from Table \ref{tab1} in Sec. \ref{multipole}) can be expressed as:
%
%\begin{widetext}
\begin{flalign}
\cF_{Co}[\mu_{Co},w_{Co}]&=
a_{0}\mu_{Co}^{2}+b_{0}w_{Co}^{2}+a_{1}\mu_{Co}^{4}\nonumber\\
&+ g\mu_{Co}\cdot w_{Co} + \dots && 
\end{flalign}
%\end{widetext}
where terms only of the order of $w_{Co}^2$ are kept.
%such as $b_{1}w_{Co}^{4}$ and higher terms can be omitted.~\cite{SIO} 
Similarly, the corresponding Landau free energy of the 
Ir sublattice ($\cF_{Ir}[\mu_{Ir},w_{Ir}]$) in terms of the two independent weakly interacting TR odd OP spin moment 
$\mu_{Ir}$ and the triakontadipole moment $w_{Ir}$, the POP of the Ir magnetic state (see Table \ref{tab1} in Sec. \ref{multipole}), can be expressed as:
\begin{flalign} 
\cF_{Ir}[\mu_{Ir},w_{Ir}]&=
a'_{0}\mu_{Ir}^{2}+b'_{0}w_{Ir}^{2}+b'_{1}w_{Ir}^{4}\nonumber\\
&+g'\mu_{Ir}\cdot w_{Ir} + \dots, &&
\end{flalign}
%\end{widetext}
keeping only terms $\sim O(\mu_{Ir}^{2})$.
%such as $a'_{1}\mu_{Ir}^{4}$ and higher terms can be omitted.~\cite{SIO} 
Thus the Landau free energy for the combined 
magnetic state can be expressed as, 
\begin{align} \label{FE}
\cF_{Co-Ir}[\mu_{Co},w_{Co},\mu_{Ir},w_{Ir}] = a_{0}\mu_{Co}^{2}+b_{0}w_{Co}^{2}+a_{1}\mu_{Co}^{4}\nonumber\\
+g\mu_{Co}\cdot w_{Co} + a'_{0}\mu_{Ir}^{2}+b'_{0}w_{Ir}^{2}\nonumber\\
+b'_{1}w_{Ir}^{4}+g'\mu_{Ir}\cdot w_{Ir} + \dots 
\end{align}

We now show that the variation of the free energy in Eq. \ref{FE} is closely related to the fixed spin moment calculations discussed above.
From the free energy expression Eq. \ref{FE}, we can determine the possible values [0, $\pm M_{Co}$] that $\mu_{Co}$ can spontaneously take using the condition (extremising $\cF_{Co-Ir}$ with respect to $w_{Co}$ and $\mu_{Co}$) $\frac{\partial\cF_{Co-Ir}}{\partial w_{Co}} = \frac{\partial\cF_{Co-Ir}}{\partial \mu_{Co}} = 0$. After some algebra we get, $M_{Co}$ =0 or $\pm\sqrt{\frac{g^2-4a_0b_0}{8b_0a_1}}$. These three points correspond to the extrema in the free energy. 
Indeed, such a behaviour is observed in Fig. \ref{fig3} (a), showing the mirror symmetry in the total energy of the system about the special point $\mu_{Co}$ = 0. While below a certain temperature, where the coefficient $a_{0}\leq 0$ the POP $\mu_{Co}$ spontaneously gets a value $M_{Co}$, the SOP $w_{Co}$ gets an induced value through the interaction term even if the coefficient $b_{0} > 0$. 

Now, focussing on the Ir sublattice, we see that below a certain temperature, where the coefficient $b'_{0}\leq 0$ the POP for the Ir sublattice $w_{Ir}$ spontaneously gets a value $w_{0} \neq 0$, determined by the condition
$\frac{\partial\cF_{Co-Ir}}{\partial w_{Ir}}=\frac{\partial\cF_{Co-Ir}}{\partial \mu_{Ir}}=0$. 
At this energy minimum the SOP $\mu_{Ir}$ gets an induced value through the interaction term although the coefficient $a'_{0}>0$. The value of this induced moment can be obtained from the condition $\frac{\partial\cF_{Co-Ir}}{\partial \mu_{Ir}}=0, \Rightarrow m_0 = -{g^\prime} w_0/{2a^\prime_{0}}$, once we know $w_{0}$. 
%This induced moment $m_{0}$ can be estimated through
%This induced moment becomes %$m_{0}$ can be estimated through
%$m_0= -{g^\prime} w_0/{2a^\prime_{0}}$. %(\comment{Please check this relation}).
%\begin{align} 
%\frac{\partial\cF}{\partial m}%=\cF_{m}+\cF_{w}+\cF_\mathrm{int}
%=2a_{0}m+g w_{0}=0
%\quad \Rightarrow\quad m_{0}= -\frac{g}{2a_{0}} w_{0}
%\end{align}

We can get the free energy variation with $\mu_{Ir}$, viz., $\cF_{Co-Ir}(\mu_{Ir})$,  around this minimum when $w_{Ir}$ is simultaneously optimized, i.e.,~under the constrain $\frac{\partial\cF_{Co-Ir}}{\partial w_{Ir}} = 0$. 
For a weak interaction the POP is essentially constant $w_{Ir}\approx w_{0}$ as we vary $\mu_{Ir}$. Further, $\mu_{Co}$ and $w_{Co}$ being non-interacting with Ir, these OPs are not expected to change. Under this condition, the free energy given in Eq. \ref{FE} becomes,
\begin{eqnarray} \nonumber \label{fixspin2}
\cF_{Co-Ir}(m_{Ir}) &=& a'_{0}\mu_{Ir}^{2}+g'\mu_{Ir}\cdot w_{Ir}\\ 
&\approx & a'_{0}(\mu_{Ir}-m_{0})^{2}+\mathrm{const},
\end{eqnarray}
where we have used the value of the induced moment $m_0= -{g^\prime} w_0/{2a^\prime_{0}}$, as obtained earlier. 
 From the above expression of the free energy, it is clear that there exist two degenerate energy minima at $\{m_{0},w_{0}\}$ and $\{-m_{0},-w_{0}\}$, for the TR symmetric free energy, i.e., $\cF_{Co-Ir}[\mu_{Ir},w_{Ir}]=\cF_{Co-Ir}[-\mu_{Ir},-w_{Ir}]$. Hence one can deduce that there have to be two independent parabolas of Eq.~(\ref{fixspin2}) centered at $\pm m_{0}$, respectively. One of these two parabolic behaviours is near about obtained in 
Fig. \ref{fig3} (c). 
%
%{\color{blue} [Shreemoyee di: Can you please add some discussions based on Landau free energy theory, explaining the above point.]} 
%It is not the dominant POP of the system like the Co spin moment is and thus the energetics of the system is not a symmetric function of it as it is for the Co moment. Interestingly as we had seen from the analysis presented in Fig.~\ref{fig3}, the SOP Ir spin moment does not depend on the POP Co moment for its existence.

 Note that the free energy model (\ref{FE}) considered in the above analysis igonres the non-collinear magnetic structure of LCIO. The appropriate model for the non-collinear magnetic structure, reflecting the underlying lattice symmetry, should be more complex \cite{Harris}. Nevertheless, the simple model, discussed above, provides useful insight into the behavior of the free energy with the variation of the primary and the secondary order parameters that help to interpret the fixed spin moment results.
We now proceed to detect the POP using the magnetic multipole analysis.
%{\color{red} I think putting that figure will raise more questions than solving them and I don't have enough data to defend all questions. Maybe you can mention the crux that w$^{415}_{Co}$ is not significantly affected by variation of corresponding spin moment.
%{\color{blue} [Shreemoyee di: Will it be also useful to add the figure that you have sent me (along with some discussions), the $w^{415}$ variation with the Co-spin moment to clarify that $w^{415}$ is also more or less unaffected.]}

\subsection{Magnetic Multipole Analysis} \label{multipole}

In this section, using the magnetic multipole analysis we will determine the POP responsible for the magnetism at the Co and Ir site. In order to do so, the computed stable magnetic states for both LCIO and LZIO are expanded in terms of the multipole tensors $w^{kpr}$, a brief description of which is given in the Appendix. 
%{\color{blue} [Shreemoyee di: What is the corresponding expression?]}. 
In terms of the said multipole moments the on-site exchange 
energy for a state (E$_x$) can be expanded as~\cite{Cricchio}:
\begin{equation}
E_x = \frac{1}{2}\sum_{kpr}K_{kpr}w^{kpr}\cdot w^{kpr},
\end{equation}
where $K_{kpr}$ are the corresponding exchange energy coefficients~\cite{Bultmark-Mult}. The exchange energy associated with a multipole is a measure of its stability. The hierarchy of the 
significant multipoles associated with Co atom and Ir atom
are listed in Table \ref{tab1}.
Note that only the multipoles which have odd TR symmetry are listed in the Table as they are the magnetic multipoles, responsible for the formation of the stable magnetic states in LCIO and LZIO.

\begin{table*}
\centering
\caption{Detail of the TR odd tensor moments in LCIO and LZIO. In LCIO, tensor moments are listed for both Co (3$d$ constituent) and Ir (5$d$ constituent) sites. The listed magnitude of the polarization $\pi^{kpr}$ and the corresponding onsite exchange energies $E_\mathrm{x}^{kpr}$
clearly show that while the spin moment is the POP for the magnetism in Co, the POP for the Ir magnetism is triakontadipole both in LCIO and LZIO.
}
\begin{tabular} {|c|c|c|c|c|c|c|c|c|c|c|}
\hline  
Name of & $kpr$ & \multicolumn{6}{|c|}{LCIO} & \multicolumn{3}{|c|}{LZIO} \\
  multipole &   & \multicolumn{3}{|c|}{Tensor moments at the} & \multicolumn{3}{|c|}{Tensor moments at the} & \multicolumn{3}{|c|}{Tensor moments at the}\\
            &   & \multicolumn{3}{|c|}{Co site}    &  \multicolumn{3}{|c|}{Ir site} & \multicolumn{3}{|c|}{Ir site}\\
    &   & $E_\mathrm{x}^{kpr}$  & $w_{kpr}$ & $\pi^{kpr}$ & $E_\mathrm{x}^{kpr}$  & $w_{kpr}$ & $\pi^{kpr}$ & $E_\mathrm{x}^{kpr}$  & $w_{kpr}$ & $\pi^{kpr}$\\
 &  & meV/atom          & $\mu_B$/atom &   &  meV/atom          & $\mu_B$/atom &         &  meV/atom          & $\mu_B$/atom &         \\  
\hline
Spin &011 & $-1423.84$&2.37 &0.254  & $-0.71$ & 0.065 &   0.0002  & -2.77 & 0.091 & 0.0003 \\
Moment&  &         &      &    &         &      &   
 &         &      & \\
Orbital&101 & $-14.90$ &0.224&0.005 & $-3.11$ & 0.149& 0.002  & -7.45  & 0.325 & 0.002\\
Moment&  &         &      &    &         &      &   
 &         &     &  \\
% Octupole&213 & $-9.05$ &0.265&0.003 & $-32.72$ & 0.703& 0.017 & -69.33 & 0.72 & 0.018\\
% Moment&  &         &      &    &         &      &   &         &      & \\
Octupole &  413 & -89.92 &  4.12 &  0.024 & -0.45  &  0.397 &  0.0002 & 0.74 &  0.36 & 0.0002\\
 Moment &  &         &      &    &         &      &   
 &         &      & \\
 Hexadecapole&414 & $-166.29$ &3.71& 0.045 & $-0.84$ & 0.36 &0.0004 & -0.56 & 0.21 & 0.0001\\
 Moment&  &         &      &    &         &      &   
 &         &      & \\
 Triakontadipole&415 & $-512.69$&8.80& 0.138& $-58.75$ & 4.07 &0.026 & -120.00 & 4.11 & 0.026\\
 \hline
%&total&&& 0.64 \\
%\hline
\end{tabular}
\label{tab1}
\end{table*}

{\it Concepts of Polarization:} It is noteworthy that in Table \ref{tab1}, the quantity $\pi^{kpr}$, known as ``polarization", is useful to compare the magnitudes of different multipole tensors instead of ${w}^{kpr}$, the former being a normalization independent quantity. To understand this fact, let us start from the condition that the density matrix $D$ satisfies the relation Tr$D$~$\geq$~Tr$D^2$. 
%for $D$ to represent a physical occupation matrix for the itinerant system. This follows from the fact that eigen values of $D$ are real and between 0 and 1. 
This condition translates to the criteria~\cite{Cricchio} 
\begin{flalign}\nonumber
&\sum_t\sum_{kpr\neq 000}2(2\ell+1)(2k+1)(2p+1)(2r+1) |N_{kpr\ell}\,{w}^{kpr}_{t}|^2 \\ 
&=\sum_t\sum_{kpr\neq 000}\pi_t^{kpr} \leq nn_h,  
%n_h &=2(2\ell+1)-n \nonumber
\end{flalign}  
%\comment{[In the above eqn I have replaced $\pi^{kpr}$ by $\pi_t^{kpr}$, and add an additional sum, please check whether it is correct]}
where $n_h=2(2\ell+1)-n$, $n$ being the number of electrons in the $d$ orbital ($l = 2$) and $n_h$ is the number of holes. The entity
\begin{align} 
\pi^{kpr}_{t}=2(2\ell+1)(2k+1)(2p+1)(2r+1) |N_{kpr\ell}\,{w}^{kpr}_{t}|^2\,
\end{align}
is the t$^{th}$ element of the ``polarization" $\pi^{kpr}$. Since $N_{kpr\ell}$ is the normalization factor for the multipole moment ${w}^{kpr}$, the construction of the moments involve multiplication by the factor $N_{kpr\ell}^{-1}$. However, as the expression for the ``polarization" involves the product $N_{kpr\ell}\,{w}^{kpr}$, $\pi^{kpr}$ is independent of the normalization constant $N_{kpr\ell}$, and therefore is useful to compare the strengths of the different multipole tensors. The multipole moments, obtained by acting the multipole operator on the density matrix, represent the various polarization channels of the density occupation matrix, viz., 
spin polarization, orbital polarization, octupole polarization, and so on. 
Only ${w}^{000}$ correspond to the total charge of the system. Therefore, if all multipole moments other than ${w}^{000}$ are 0, then the system is unpolarized. This is why the quantity $\pi^{kpr}$ is called ``polarization"  which is nothing but the trace of square of the density matrix, 
and involves the square of all multipole moments other than ${w}^{000}$. So, all contributions, excluding $kpr=000$, add up to a total polarization 
\begin{align} 
\pi^\mathrm{tot}=\sum_t\sum_{kpr\neq 000}\pi^{kpr}_{t} .
\end{align} 

%------------------------------------------
%In order to be able to compare the magnitude of different multipole tensors, a normalization independent quantity has been introduced, the polarization 
%\begin{align} 
%\pi^{kpr}_{t}=2(2\ell+1)(2k+1)(2p+1)(2r+1) |N_{kpr\ell}\,{w}^{kpr}_{t}|^2\,.
%\end{align}
%Here `t' denotes the component of the tensor of rank `r'. The normalization constant for tensor moment 
%${w}^{kpr}$ (corresponding to orbital quantum number l) is given as $N_{kpr\ell}$ (details given in 
%Ref.\cite{Cricchio,Laan}). 
%All contributions, excluding $kpr=000$,
%add up to a total polarization 
%\begin{align} 
%\pi^\mathrm{tot}=\sum_{kprt}\pi^{kpr}_{t}\,,
%\end{align} 
%which is constrained by the inequality 
%\begin{align} 
%\pi^\mathrm{tot}\leq (10-n_d)n_d\,,\label{ineq}
%\end{align} 
%where $n_d$ is the occupation number of the $d$-shell.

%{\color{blue} ------------------------------------------}

{\it LCIO:} As we can see from Table \ref{tab1} the exchange energy $E_\mathrm{x}^{kpr}$ and the polarization $\pi^{kpr}$ associated with the magnetic multipole corresponding to the Co-spin moment is the highest. This reconfirms our earlier detection of the Co spin moment as the POP for the Co magnetic state. The triakontadipole 
moment which is formed by coupling of spin and orbital moment is the next significant magnetic OP. 
On the other hand, for the Ir atom, as is 
evident from Table \ref{tab1}, the most significant multipole is the triakontadipole 
moment. The exchange energy and the polarization is highest for the said moment amongst all 
the magnetic multipole moments corresponding to the Ir magnetic state. So, the mentioned 
magnetic state is primarily formed by the triakontadipole moment, i.e., it is the POP for the 
state. Note that the magnitude of the spin moment $\mu_s$ is only 0.07$\mu_B$/Ir (See Table \ref{tab1}), as it is a SOP induced by the POP triakontadipole moment. As a result, the total moment at the Ir site also significantly deviates from 1 $\mu_B$, the value predicted within the $j_\mathrm{eff} = 1/2$ model \cite{jeff}. We note that the orbital to spin moment ratio in LCIO is $\sim$ 2.1, similar to the value $\frac{\mu_l}{\mu_s}$~$\sim$~2, predicted from the $j_\mathrm{eff} = 1/2$ model~\cite{jeff}.
%Also we can see from Table \ref{tab1} that the ratio of Ir orbital moment 
%($\mu_l$ = 0.15 $\mu_B$/atom) and spin moment ($\mu_s$=0.07$\mu_B$/atom) is $\sim$ 2.1, alligned with the ratio $\frac{\mu_l}{\mu_s}$~$\sim$~2, predicted within the $j_\mathrm{eff} = 1/2$ model~\cite{jeff} for the Ir atom. 

 Further, to gain insights into the role of the SOC in LCIO we turned off the SOC, performing a  DFT+U only study. In this case, the 
values of the most significant tensor moments of the said system are listed in Table \ref{tab2}.
\begin{table}
\centering
\caption{Multipole moments at the Co and Ir sites in LCIO in the absence of SOC.} 
\begin{tabular} {|l|l|c|c|c|c|}
\hline 
Element& Name of &  $kpr$ & $w_{kpr}$  & $E_\mathrm{X}^{kpr}$ & $\pi^{kpr}$\\
       & Multipole &      &            & & \\
         &           &     & $\mu_B$/atom   & meV/atom & \\
\hline
 & Spin Moment        & 011 &  2.53       & -3229.36  &  0.29  \\
Co & Orbital Moment     & 101 &0.00        & 0.00    & 0.00\\
 &Triakontadipole          & 415 & 6.45     & -551.22 & 0.07    \\ 
\hline
&  Spin Moment  & 011 & 0.46 & -72.33  & 0.01\\
Ir&  Orbital Moment & 101 & 0.00  & 0.00 & 0.00\\
&  Triakontadipole & 415 & 2.59  & -47.56 & 0.01\\
\hline
\end{tabular}
\label{tab2}
\end{table}
 As we can see from Table \ref{tab2} if the effect of SOC is not considered, the spin 
moment becomes the dominant magnetic multipole for the Ir atom  and its magnitude increases to 0.46 $\mu_B$/Ir. Thus SOC plays the major role in promoting the higher order multipole 
moment, the triakontadipole moment, to be the POP to drive the magnetic state. Very interestingly, in this situation the Co 
 and Ir moments are ferromagnetically coupled. This, further, points out the importance of the SOC in driving the non-collinear magnetic structure in the system.

{\it LZIO:} Similar to LCIO, the listed hierarchy of the tensor moments [see Table \ref{tab1}] at the Ir site in LZIO clearly shows that   
the triakontadipole moment is the POP responsible for the Ir magnetic 
state with the highest exchange energy and polarization associated with it. 
  We note that the orbital to spin moment ratio $\frac{\mu_l}{\mu_s}$
 in LZIO is as large as 3, similar to the other reported $d^5$ iridates \cite{Marco}. The larger $\frac{\mu_l}{\mu_s}$ ratio as well as the larger exchange energy corresponding to the triakontadipole moment at the Ir site in LZIO as compared to LCIO may be attributed to the enhanced effective SOC in the former compound due to the associated relatively smaller band width and weaker noncubic crystal field, as discussed earlier. 
In spite of the enhanced effective SOC in LZIO, the multipole analysis clearly shows that the Co magnetic states (with the spin moment as the POP) hardly affects the Ir magnetic state (having triakontadipole as the POP).

\subsection{Tight-binding Analysis}

\begin{figure}[h]
\centering
%\vspace{2 cm}
%\resizebox{12cm}{!}
{\includegraphics[width=\columnwidth]{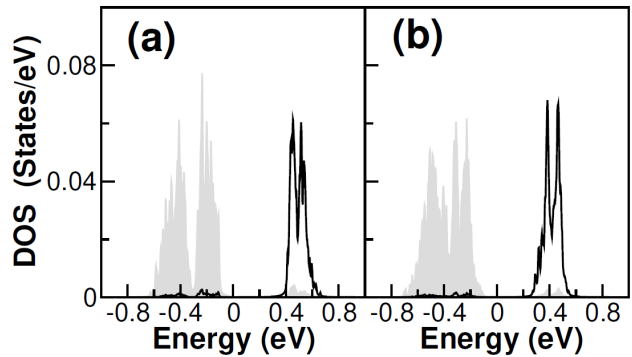}}
\caption{The results of the tight-binding model (\ref{TBM}) for Ir-$t_{2g}$ orbitals. The Ir-$t_{2g}$ density of states, projected onto $j_{\rm eff}$ = 1/2 (solid line) and $j_{\rm eff}$ = 3/2 (shaded region) (a) including and (b) excluding the effect of Co-$d$ states in LCIO. The very small change in the mixing between $j_{\rm eff}$ = 1/2 and $j_{\rm eff}$ = 3/2 states in both the cases indicate that Co-$d$ states hardly affect the Ir $d$ states.}
\label{fig4}
\end{figure}

The impact of the Co-$d$ states on the Ir-$d$ states are, further, studied within a tight-binding (TB) analysis. We have constructed a low-energy Ir-$t_{2g}$ only TB model Hamiltonian in presence of SOC:
\begin{eqnarray} \label{TBM}\nonumber
 \mathcal{H}& =  & \mathcal{H}_{kin} + \mathcal{H}_{soc}\\\nonumber 
  & = & \sum_{i\mu\sigma, j\nu\sigma} t^{\mu \nu}_{ij} c_{i\mu \sigma}^\dagger c_{j\nu \sigma} 
 + \frac{\lambda}{2} \sum_{i\eta} \sum_{\mu\sigma,\nu\sigma^\prime} c_{i\mu \sigma}^\dagger 
L^\eta_{\mu\nu}   \Sigma^\eta_{\sigma\sigma^\prime} c_{i\nu \sigma^\prime},\\
\end{eqnarray}
where $i$-$\mu$-$\sigma$ represent the site-orbital-spin indices, $\eta$ denotes the three cartesian components, 
$\lambda$ is the SOC constant that couples the orbital angular momentum $\vec L$ and the spin $\vec s = \frac{1}{2} \vec \Sigma$ of the electron, and $t^{\mu \nu}_{ij}$ represent the hopping between Ir-$t_{2g}$ orbitals, extracted from the DFT calculation using NMTO downfolding method~\cite{AndersenJepsen}. Hopping upto six nearest neighbors are considered in the model. The NMTO downfolding calculation is carried out first keeping only the Ir-$t_{2g}$ states and then keeping both Ir-$t_{2g}$ 
states and Co-$d$ states in the basis. Therefore, we have the hopping parameters between the Ir-$t_{2g}$ orbitals including and excluding the effect of Co-$d$ states respectively. The densities of states for the Hamiltonian (\ref{TBM}) is calculated with the Lorentzian-broadening for these two cases and are projected on to the $j_{\rm eff}$ = 3/2 and $j_{\rm eff}$ = 1/2 states of Ir atom as shown in 
Fig~\ref{fig4}. As it is clear from Fig~\ref{fig4}, the mixing between $j_{\rm eff}$ = 3/2 and $j_{\rm eff}$ = 1/2 states are almost unaffected with the inclusion of the effect of Co-$d$ states. This indicates that the Co-$d$ states do not alter the  spin-orbit entangled Ir-$j_{\rm eff}$ = 1/2 states in agreement with the multipole analysis.
%finding of the higher order multipole $w^{415}$ as the POP for Ir in presence of the Co spin magnetic moments, which happens to be the POP for the magnetic state of Co.

\section{Summary}
In summary, a comprehensive study of the magnetism in the double perovskite compound LCIO, having both magnetic 3$d$ (Co) and 5$d$ (Ir) TM atoms is performed in the present work using fixed spin moment and magnetic multipole analysis. In order to comparitively decipher the effect of the magnetic 3$d$ TM species on the TR odd state of Ir, the isostructural double perovskite iridate LZIO, having a non-magnetic 3$d$ and the same 5$d$ (Ir) TM species, has also been studied. From the comparisons of the noncubic crystal fields arising from the distortions 
in the respective crystal structures, band-width, and Ir-O covalency we had inferred that the effect of SOC would be more predominant in LZIO than in LCIO. %The mentioned multipole analysis confirms our inference. 

A detailed fixed spin moment analysis in the light of the extended Landau free energy analysis and the multipole analysis provide useful insights into the magnetic order parameters responsible for magnetism in the 3$d$ and 5$d$ species in LCIO. 
From the fixed spin moment analysis, we detect that the Co-spin moment is the dominant POP for the magnetic state of Co. This is, however, not the case for the 5$d$ species Ir as the Ir spin moment is merely a SOP. 
%In fact the ratio of orbital to spin moment indicates that the Ir 5d state in \lacoo~has the symmetry of the $j_{eff}$=1/2 state for strong SOC. 
Explicit magnetic multipole analysis shows that the Ir magnetic state is primarily driven by the spin-orbit coupled higher order magnetic multipole moment, the triakontadipole moment. Comparison with the isostructural LZIO, although shows a larger exchange energy associated with the triakontadipole moment in LZIO, indicating a stronger effective SOC in this material compared to LCIO, the identification of the triakontadipole moment as the POP at the Ir site in LCIO clarifies the existence of the spin-orbit entangled Ir-$d$ states even in presence of Co-$d$ states. This is, further, corroborated by the tight-binding analysis, which confirms negligible influence of the Co-$d$ states on the Ir-$j_{\rm eff}$ = 1/2 states in LCIO. 
 The multipole ordering in LCIO may manifest itself in the magnetic susceptibility
measurements, in which a change in the slope of the inverse susceptibility may be observed.
Interestingly, such a behavior is observed for LCIO \cite{Lee}.
Our work, therefore, calls for further mesurements in this direction for the experimental confirmation of the computed mutipole ordering in LCIO. Furthermore, 
%the magnetic multipoles may also be probed in neutron diffraction experiments.
the computed hierarchy of the relevant multipoles in LCIO may also be captured in neutron diffraction experiments by determining their relative intensity contributions to the magnetic neutron Bragg spots \cite{Lovesey}.  Finally, LCIO being a representative of the 3$d$-5$d$ double perovskite family, the results of the present work provide a guidance to understand the magnetism in this broad family of materials.
%
%Our study therefore motivates more study based on magnetism of the 3$d$-5$d$ systems, which are potential to host two different magneti species, the individual magnetic states of which are driven by two different magnetic order parameters. {\color{red} Neutron diffraction experiments might capture  for a system in a certain magnetic state.~ [] This hierarchy is determined } {\color{blue} [Shreemoyee di: Is there any experimental tool to detect the higher order multipoles, eg., triakontadipole? Can they have possible applications? Such informations might provide the relevance of the present work.]} 
%
\section{acknowledgement}
The authors thank Indra Dasgupta for useful discussions. SG would like to thank Lars Nordstr\"om and Oscar 
Gr{\aa}n\"as for valuable suggestions and guidance during the course of this work. 

{
\vspace {2 cm}
\centering
{\bf {\center APPENDIX: MULTIPOLE ANALYSIS}}
}
\vspace {.2 cm}

The Hartree-Fock term (E$_{HF}$) in the LDA+U method is given by,
\begin{equation}
E_{HF}=\frac{1}{2}\sum_{abcd}\rho_{ac}[\langle ab|q|cd \rangle - \langle ab|q|dc \rangle]\rho_{bd},
\end{equation}
where $\rho_{ac}$ is an element of the density matrix $D$, discussed previously, that acts as an occupation matrix and a, b, c and d are single-electron states. The interaction term 
$\langle ab|q|cd \rangle=\int \psi_a^\dagger(1)\psi_b^\dagger(2)q(r_{12})\psi_c(1)\psi_d(2)d(1)d(2)$ signifies interaction between electrons (1) and (2). The wave function $\psi$ is a single-electron wave 
function for a state with magnetic and spin quantum numbers $m$ and $s$ respectively. The interaction is taken to be of 
the form of a screened Coulomb interaction in the present study. So, $q(r_1,r_2)=\frac{e^{-\lambda r_{12}}}{r_{12}}$, where $\lambda$ is the screening parameter. The interaction term can be 
expressed in terms of spherical harmonics. Using Wigner 3$j$, 6$j$ and 9$j$ symbols~\cite{Condon,Racah,Bultmark-Mult} and Slater integrals~\cite{Slater} the Hartree-Fock energy $E_{HF}$ can be expressed in terms of the irreducible spherical tensors known as multipole tensors, a detail description of which can be found in Ref.~\onlinecite{Bultmark-Mult}. 
%Now, the Hartree-Fock Energy has a spin independent part, the Hartree term and a spin-dependent part, the Exchange term.

\end{document}